\documentclass[
  reprint,
  aps,
  pre,
  amsmath,
  amssymb
]{revtex4-2}

\usepackage{graphicx}

\begin{document}

\title{Balancing innovation and assimilation in research communities}

\author{Bill Nunn$^\star$, Maria Horner$^\dag$, Marcel Ortgiese$^\star$, Tim Rogers$^\star$}
\affiliation{$^\dag$Imperial College London, South Kensington, London SW7 2AZ, United Kingdom, $^\star$Department of Mathematical Sciences, University of Bath, Bath, BA2 7AY, United Kingdom}

\email{Bill Nunn: wn256@bath.ac.uk}


\begin{abstract}

It has been statistically observed that the speed at which a corpus of knowledge advances does not scale linearly with quantities like the number of active researchers or number of research papers published in a given field. Furthermore, as a body of knowledge grows, individual researchers must somehow strike the right balance between generating new knowledge through innovation and assimilating knowledge generated by others. Here, we propose and analyse a pair of interacting particle system models representing some stylised features of the advancement of knowledge in a research community, captured as a stochastic travelling wave in knowledge space. Both particle systems exhibit a diminishing return in the knowledge advancement speed as research communities grow larger, and suggest that researchers should spend more time on assimilation of knowledge than innovation as their research community grows.
\end{abstract}

\maketitle

\section{Introduction}\label{sec:int}

\begin{figure*}
    \centering
    \includegraphics[width=0.9\linewidth]{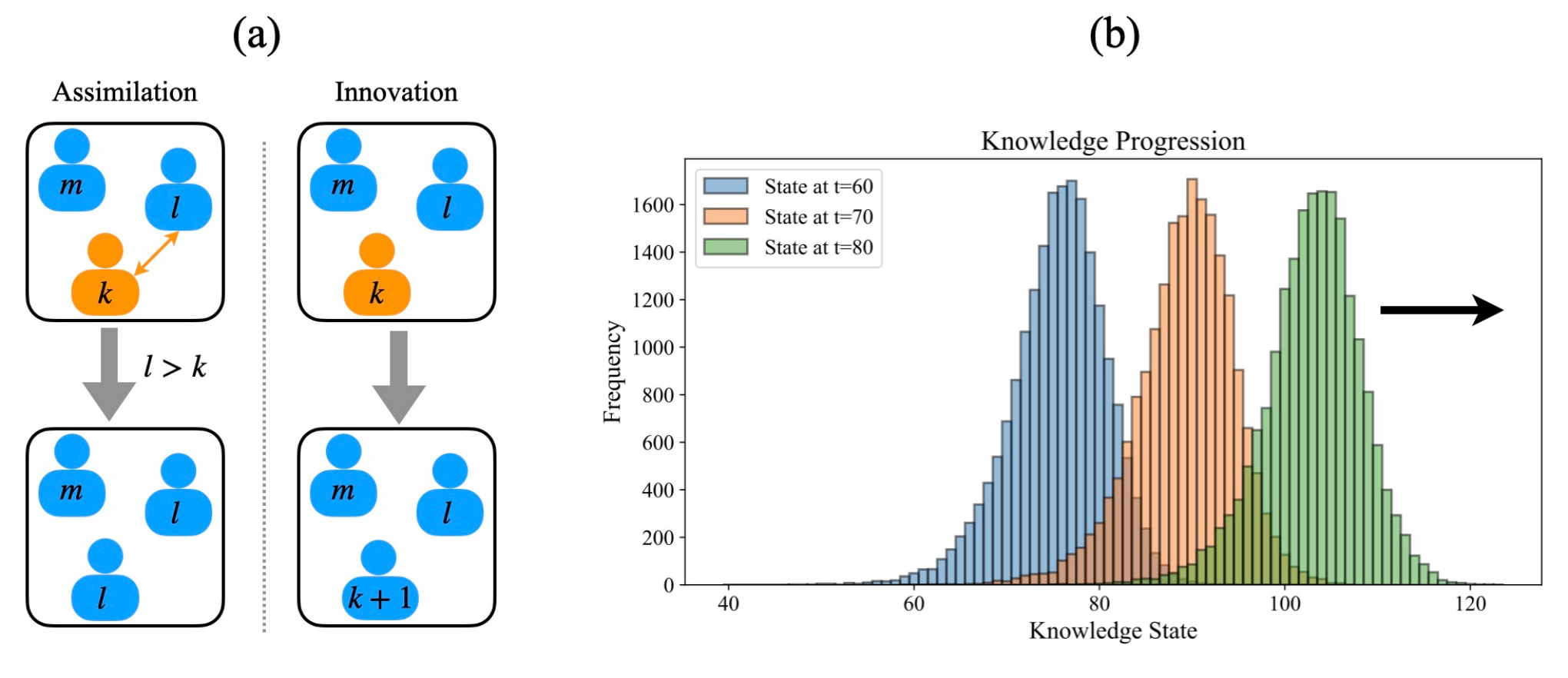}
    \caption{Panel (a) shows a schematic for how the knowledge states in the community of researchers update. The researcher in orange has just had their Poisson clock ring. For the assimilation action they choose to interact with the researcher with knowledge $l$. Panel (b) plots the number of researchers with each knowledge state as time progresses for the low information case, with $N=10,000$ and $q=1/2$. Recall the $q$ allocation approximates the proportion of the time each researcher spends on innovation. We see the knowledge states of the researchers progress as a non-diffusive stochastic travelling wave, travelling rightwards.}
    \label{fig:schem}
\end{figure*}

The `science of science'~\cite{Fortunato2018} has produced many intriguing statistical findings about large research communities. Perhaps the most famous observation is Price's Law~\cite{Price1963}, which states that half of the research output of any community derives from the square root of the total number of contributors. Mathematical models are not common in the science of science, and mechanistic issues are often considered sociologically instead. Notable exceptions include usage of epidemiological models to predict the number of researchers in a given field~\cite{Goffman1966} and the rate of adoption of a particular scientific idea by a research community~\cite{Bettencourt2006}. Preferential attachment models also offer mechanistic insight into citation statistics~\cite{Wang2013, Masucci2011}, recovering the empirically observed distributions of citation counts. Mechanistic understanding of the speed in which scientific knowledge advances has proven more elusive however. 

Quantity of papers produced is not a good proxy for the speed of knowledge advancement. While the number of papers published each year has grown exponentially over the past century~\cite{Bornmann2021}, information-theoretic measures of knowledge derived from citation networks indicate sub-exponential growth~\cite{Kang2024}. Furthermore, papers have been shown to become less \emph{disruptive} over time~\cite{Park2023}. A paper is deemed disruptive if the papers which cite it do not frequently cite the papers it cites. The intuition is that such a citation pattern would be observed in papers which make prior work redundant. The (controversial) suggestion is that most modern papers contribute less original knowledge than their older counterparts did. Given the lack of clear metric for knowledge advancement, the door is left open for a more abstract mathematical enquiry.

We will consider a family of interacting particle systems (IPSs) called Knowledge Advancement Processes (KAPs). Each KAP is a continuous time stochastic process taking place in a community of $N$ researchers. Knowledge of the researchers is represented by an infinity of ordered states. IPSs with many ordered states have been considered in a wide variety of domains: multistage epidemic models \cite{Stanoev2014, Karrer2011}; rigorous probabilistic inquiry of multi-type contact processes \cite{Neuhauser1992, Foxall2016, Seiler2026}; and, close to our topic, the rise and fall of scientific paradigms and trends \cite{Mirshahvalad2014, Bornholdt2011, Lee2024}. The IPSs most similar to KAPs are those considered by Sontag \cite{Sontag2023,sontag2022misinformation,sontag2024dynamics}, who generalised the `awareness' model introduced by Funk \cite{Funk2009}. Funk's original model tracked the awareness of a receiving individual by counting the number of people a piece of information had passed through en route from an information source to the receiver. 

Each researcher in the community has an integer valued knowledge state corresponding to their knowledge of a specific branch of some research field. Every researcher is equipped with an associated unit rate Poisson clock, assumed mutually independent. When a researcher's clock rings they perform one of two actions: assimilation or innovation of knowledge. If the researcher performs an assimilation action another researcher in the community is chosen, and the assimilator will copy their knowledge state if it is higher. 
Alternatively, if the original researcher innovates, they spontaneously increase their knowledge state by one. 
Note that the knowledge state a researcher reaches via an innovation action might not be a new across the whole community, just new to the individual. See Figure \ref{fig:schem} for visual illustration of these events in the community of researchers.

Throughout this article we consider two KAPs, both well behaved and Markovian. We call these the high information, and the low information KAPs respectively.
\textbf{High Information (HI):}
Researchers always have full information about the knowledge state of all other researchers in the community. Upon a clock ring, a researcher innovates when (jointly) most knowledgable amongst all researchers, and assimilates knowledge from a researcher with highest knowledge otherwise.\\
\textbf{Low Information (LI):}
Researchers are persistently unaware of \emph{all} knowledge states, including their own. In the LI case we assert that each researcher randomises between assimilating and innovating. To this end we assign every researcher the same allocation $q \in (0,1)$, which gives the probability of innovating when their Poisson clock rings. One can interpret $q$ as the proportion of time researchers spends on trying to innovate. Each time a researcher assimilates they choose a member of the community uniformly at random.

The knowledge states of both processes progress as non-diffusive stochastic travelling waves; see Figure \ref{fig:schem}(b) for an illustration of the LI case. The expected speed of the stochastic travelling wave offers a proxy for the speed of knowledge advancement. To be explicit, expected speed of the stochastic travelling wave refers to the long term average speed of the wave once the wave cross section is close to `equilibrium'. To estimate the expected speed by simulation therefore, all researchers are assigned knowledge state $0$ at time zero, and we take two times $t_1 \ll t_2$. Time $t_1$ is chosen large enough so that the wave cross section is in equilibrium (and there is no memory of the starting state). We then calculate the average of the knowledge states at times $t_1$ and $t_2$ and divide the difference of the averages by $t_2-t_1$ to estimate the expected speed.

The HI and LI cases are just two possibilities, and any other specification for how each researcher decides between assimilating and innovating, and the researcher assimilated from is chosen, yields a new KAP. We remark that the HI and LI processes are extremal within the family of all KAPs. In particular, for any process in the family of KAPs the knowledge advancement speed cannot exceed that of the HI process. Moreover, the knowledge advancement speed of the LI process can always be attained regardless of whether or not there is more information available for researchers to make decisions between assimilating and innovating.

Finding estimates for the speed of the knowledge advancement in the HI and LI case therefore bounds the possible range of behaviours of the full family of KAPs. Additionally the LI process allows us to more overtly explore how the balance between innovating and assimilating (the value of $q$) impacts the speed of knowledge advancement.

The paper proceeds as follows. In Section \ref{sec:hi} we derive the expected advancement speed of the HI process on a community of $N$ researchers (and thus an upper bound for the advancement speed of all KAPs). In Section \ref{sec:li} we obtain speed estimates for the LI case, finding a finer upper bound on the LI speed in \ref{sub:tw} which depends on $q$. The exact expected speed for a community of two researchers is calculated in \ref{sub:2node}.

\section{High Information Case}\label{sec:hi}
\begin{figure}
    \centering
    \includegraphics[width=1.0\linewidth]{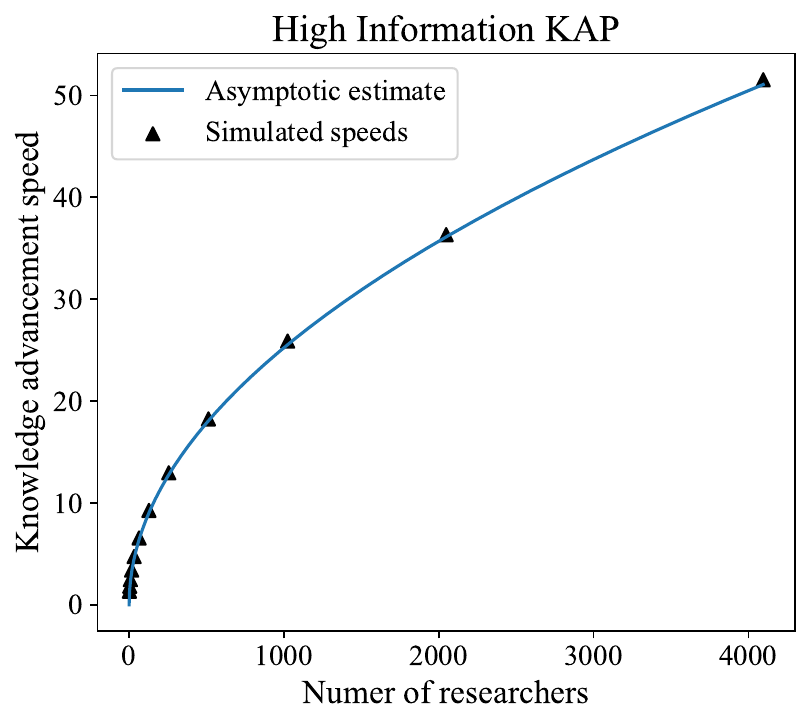}
    \caption{The knowledge advancement speed for the High Information process for a community of $N$ researchers.}
    \label{fig:hi}
\end{figure}
We show analytically that there is a diminishing return in the knowledge advancement speed as the size of the research community is increased. In particular, for the HI process on $N$ researchers the knowledge advancement speed is shown to scale as $\sqrt{N}$. Sublinear growth in $N$ is perhaps unexpected given the simplicity of the HI process. Economists prefer their metrics to be expressed per capita, and the HI process exhibits knowledge advancement per capita decreasing as the research population size increases. This qualitatively matches the behaviour of related economic metrics: the research population has grown over the 20th century but per capita research productivity steadily decreased, see \cite{Bloom2020} for example. 

Bloom \cite{Bloom2020} goes on to argue that the drop in per capita research productivity implies that ideas are becoming harder to find, and perhaps that (lots of the) low hanging fruit has been plucked \cite{Cowen2012}. Our analysis of the HI process offers at least one confounding consideration: KAPs incur a diminishing return in the knowledge advancement speed as population size increases because researchers have to assimilate more often to keep up with the progress of their community. Furthermore, presence of a diminishing return in advancement speed is in apparent agreement with recent empirical observations of Chu and Evans \cite{Chu2021}. Chu and Evans find that progress is slower per capita in fields with more researchers, even in equally mature fields.

We refer to the researchers with joint largest knowledge as being at the \emph{front} of the travelling wave. The future evolution of the HI process is fully characterised by the number of researchers at the front. To obtain the expected speed at which the front of the HI process progresses we derive the expected time taken to transition from having exactly one researcher at the front with knowledge state $k$ to having exactly one researcher with knowledge state $k+1$. The way the system transitions from the former state to the latter is by researchers behind the front jumping to the front via assimilation actions until the Poisson clock of one of the front running researchers rings and they innovate. Since actions across the whole community occur at rate $N$, to get the expected speed we need only find the distribution of the number of (community wide) actions before some researcher at the front makes an innovation action.

Let $p(n)$ denote the probability that exactly $n$ actions are required to bring the system from a single researcher at the front with knowledge $k$ to a single researcher at he front with knowledge $k+1$. Moreover, let $T$ (a random variable) denote the time required to transition between these two states. We define $r(n) := 1 - \sum_{i=1}^{n} p(i)$ to be the probability that more than $n$ actions are required, with $r(0) := 1$. It is clear that 
\begin{equation}
    r(n) = (1-\frac{n}{N})\cdot r(n-1)
\end{equation}
because after $n-1$ actions have occurred which do not progress the front there are exactly $n$ researchers at the front, and thus the probability the next action also does not progress the front is $1-n/N$. We therefore have
\begin{equation}\label{eq:cf}
    r(n) = \prod^n_{i=1} \left(1 - \frac{i}{N}\right).
\end{equation}

Next, observe that
\begin{equation}\label{eq:et}
    \mathbb{E}[T] = \sum^{N}_{n=1} p(n) \cdot \frac{n}{N} = \frac{1}{N}\sum_{n=0}^{N-1}r(n),
\end{equation}
as $n/N$ is the expected time for $n$ actions to occur (conditioned on exactly $n$ actions being required to progress the front). We will see that \eqref{eq:cf} and \eqref{eq:et} yield
\begin{equation}\label{eq:root}
\mathbb{E}[T] \approx \sqrt{\frac{\pi}{2 N}}
\end{equation}
asymptotically, and so the speed of the front is approximately $\sqrt{2N / \pi}$. See Figure \ref{fig:hi} for a comparison of this estimate with simulated speeds of the high information stochastic travelling wave.

To obtain estimate \eqref{eq:root} we first observe that
\begin{align}
    \log r(n) &= \sum_{i=1}^n \log \left( 1-\frac{i}{N}\right)
    = - \sum_{i=1}^n \sum^{\infty}_{j=1} \frac{i^j}{jN^j}, \nonumber\\
    &\approx - \sum_{i-1}^{n} \frac{i}{N} = -\frac{n(n-1)}{2N}.
\end{align}
to leading order in $1/N$. Therefore 
\begin{equation}
    r(n) \approx \exp \left(-\frac{n^2}{2N}\right),
\end{equation}
and so approximating the sum \eqref{eq:et} by an integral, and noting that the left hand tail of the integral contributes negligible mass for large $N$ yields
\begin{align}
    \mathbb{E}[T] &= \frac{1}{N}\sum_{n=0}^{N-1}r(n), \approx \frac{1}{N}\int^N_0 \exp(-\frac{x^2}{2N}) \,dx,\nonumber\\
    &\approx \frac{1}{N}\int^{\infty}_0 \exp(-\frac{x^2}{2N}) \,dx,= \sqrt{\frac{\pi}{2 N}}.
\end{align}

The $\sqrt{N}$ speed scaling relates to Price's Law. In particular, the expected size the front the HI process reaches before progressing is of order $\sqrt{N}$. It might then be that these $\sqrt{N}$ front running researchers contribute most of the research papers before the next major innovation, as is suggested by Price's Law.

The intuitive reason that the knowledge advancement shows a diminishing return in the speed is that researchers must assimilate more often the faster the the front moves. We know that the front moves at speed $\sqrt{2N/\pi}$ and that order $N$ actions are performed per unit time across the whole community. Since innovation actions are performed iff they progress the front, as a proportion of all actions only $2/\sqrt{\pi N}$ are innovation. All other actions are assimilation and we therefore see that researchers must assimilate knowledge more often as the research population $N$ grows. 

\section{Low Information Case}\label{sec:li}

In the LI case the knowledge advancement speed depends on the fraction $q$ of time spent on innovation as opposed to assimilation. In particular we find the speed to be bounded above by a finite function $u(q)$ for communities of any size. We calculate $u(q)$ in subsection \ref{sub:tw} using techniques from the travelling wave literature. Simulations of the LI process show slower knowledge advancement speeds in smaller communities. The exact knowledge advancement speed $l(q)$ is therefore calculated for a community of two researchers in subsection \ref{sub:2node}.

\subsection{Travelling Wave Estimate} \label{sub:tw}

For the LI process the travelling wave estimate
\begin{equation}\label{eq:upest}
    u(q) = \left( W\left[\frac{1-2q}{q e}\right]\right)^{-1} (1-2q)
\end{equation}
upper bounds the knowledge advancement speed for communities of any size, where $W[\cdot]$ denotes the principle branch of the Lambert $W$ function \cite{Corless1996}. Recall that $q$ is the proportion of time each researcher spends innovating. This bound contrasts the HI case in which the knowledge advancement speed grew unboundedly (but diminishingly) as research communities grew large.

We arrive at \eqref{eq:upest} by deriving a system of ODEs which approximate the behaviour of the LI process, positing a travelling wave solution over the (discrete) knowledge space, and obtain restrictions on wave speed by considering the linearisation of the system. This approach is the discrete analogue of the standard approach to obtaining speed estimates of Fisher-KPP travelling waves \cite{Murray1989}. Previous work on speed estimates for similar particle systems makes the analogue of the knowledge space continuous, see \cite{Sontag2023, Taylor2014} for example. We found that making this continuous approximation underestimates the wave speed for large communities.

Consider a community of $N$ researchers, all with the same $q$-allocation. For large $N$ The gain-loss equation \cite{Kampen2007} for the proportion of researchers in each knowledge state in the LI case is given by
\begin{widetext}
\begin{align}\label{eq:master}
    \partial_t p_{k}(t) = \underbrace{ q \cdot p_{k-1}(t)}_{\text{Innovation to knowledge $k$}} &- \underbrace{q \cdot p_{k}(t)}_{\text{Innovation from knowledge $k$}} + \underbrace{(1-q) \cdot \sum_{m<k}p_{m}(t) p_{k}(t)}_{\text{Assimilation to knowledge $k$}} - \underbrace{(1-q) \cdot \sum_{m>k}p_{m}(t) p_{k}(t)}_{\text{Assimilation from knowledge $k$}}.
\end{align}
\end{widetext}
The variable $p_{k}(t)$ approximates the proportion of researchers that have knowledge $k$ at time $t$.

Defining the cumulative variable $u_{k} := \sum_{m\geq k} p_{m}$ and carefully transforming system \eqref{eq:master} yields
\begin{equation}\label{eq:cum}
    \partial_t u_{k} = q (u_{k-1} - u_{k}) + (1-q) u_{k}(1-u_{k}),
\end{equation}
which readers might recognise as a discrete version of the Fisher-KPP equation. The linearisation of system \eqref{eq:cum} is then simply
\begin{equation}\label{eq:lin}
    \partial_t u_k = q(u_{k-1} - u_{k}) + (1-q) u_k.
\end{equation}

We now posit the travelling wave ansatz $u_k(t) = \exp (-\lambda (k-vt))$. Substituting the travelling wave ansatz $\exp (-\lambda (k-vt))$ into the linearised system \eqref{eq:lin} we obtain the following dispersion relation for the travelling wave speed $v$ in terms of $\lambda$
\begin{equation}\label{eq:speed}
    v(\lambda) = \frac{q(e^\lambda - 1) + (1-q)}{\lambda}.
\end{equation}
As the front is pulled the speed system \eqref{eq:lin} attains is given by the minimum of \eqref{eq:speed} over all $\lambda >0$. Performing the elementary calculus  task yields estimate \eqref{eq:upest}, plotted as the solid upper curve in Figure \ref{fig:li}. We see that there is a trade-off to be made between innovating and assimilating and the fastest advancement speed occurs when both innovation and assimilation take place. The travelling wave estimate \eqref{eq:upest} upper bounds the (expectation of the) simulated speeds of the LI process in finite communities.

The simulated speeds converge to the travelling wave estimate as $N$ gets large. The convergence to the travelling wave estimate is unusually slow however, with Brunet-Derrida \cite{Brunet1997} style correction to the speed scaling like $(\ln N)^{-2}$. We can observe the slow convergence in Figure \ref{fig:li}. The most important feature of these simulated speed curves is that the location of the peak changes as the number of researchers is varied from $N=2$ up to $N=4000$. We see that when the research population gets larger the $q$-allocation which maximises knowledge advancement speed sits further to the left. In other words, to optimise knowledge advancement speed in the LI case researchers must spend a larger proportion of their time assimilating knowledge the larger their research community.

\subsection{Exact Speed for $N=2$}\label{sub:2node}

\begin{figure}
    \centering
    \includegraphics[width=\linewidth]{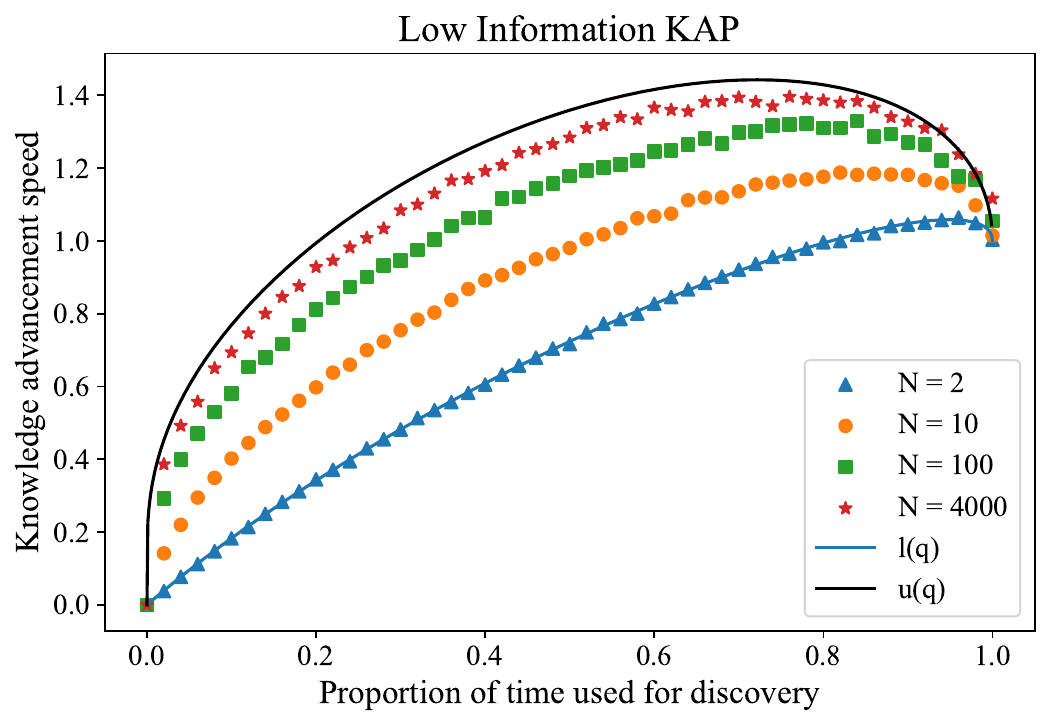}
    \caption{Knowledge advancement speed of the Low Information KAP in a community o $N$ researchers. The horizontal axis shows the proportion of time that researchers innovate. The upper and lower bounding curves are derived in Section \ref{sec:li}.}
    \label{fig:li}
\end{figure}

To obtain accurate speed estimates for small communities, more involved probabilistic arguments are required. We derive the exact expected speed for the very simplest case $N=2$. In the case of a community of two researchers we find the expected knowledge advancement speed of the LI process is 
\begin{equation}\label{eq:two}
    l(q) = q + q\sqrt{\frac{1-q}{3q+1}}.
\end{equation}
This function is plotted in Figure \ref{fig:li} showing agreement with the simulated knowledge advancement speeds for two researchers.

To arrive at \eqref{eq:two} we break the process up into times in which one researcher assimilates knowledge from the other. We can get an exact distribution for the length of time between such events, as well as the expected amount of knowledge gained via innovation in this time. Having both allows us to get at the speed.

Let us mark one of the two researchers arbitrarily. To begin the speed calculation we define a \emph{successful jump} to be the outcome that the less knowledgeable researcher performs an assimilation action, or the marked researcher performs a assimilation action when both researchers have the same knowledge state. The apparent asymmetry in this definition ensures there are exactly $G \sim \text{Geometric}(1/2)$ many assimilation actions between successful jump occurrences.

It is useful to split the process into time intervals between successful jumps for two reasons. First, at the moment these intervals start one of the researchers has just assimilated knowledge from the other and so both have identical knowledge state. Second, any assimilation action within the interval does not result in knowledge gain, otherwise it would be classed as a successful jump.

The time between successful jumps is
\begin{equation*}
    \mathcal{T} = \sum_{n=1}^{G} T_n
\end{equation*}
where $T_n \sim \text{Exponential}(2(1-q))$ are independent of each other, and the times taken to induce assimilation actions. The random variable $\mathcal{T}$ can be shown to be distributed as an exponential random variable with scale parameter $(1-q)$ by considering the moment generating function of the above sum.

Without loss of generality assume that the researchers start with identical knowledge, i.e. a successful jump occurs at time zero. Furthermore, suppose that the time until the next successful jump is $\tau$. The knowledge state of both researchers at time $\tau$ is then $\max \{K_1, K_2\}$ where $K_1$ and $K_2$ are independent identically distributed Poisson random variables with rate parameter $q\tau$. Defining the expected progress $f$ of the knowledge states after time $\tau$, conditioned on the fact that the next successful jump occurs at time $\tau$, we have
\begin{align}\label{eq:prog}
    f(\tau) &= \mathbb{E}[\max \{K_1,K_2\}] \\ &= \mathbb{E} [\frac{1}{2}(K_1 + K_2 + \lVert K_1-K_2 \rVert)] \\
    &= q\tau + \frac{1}{2} \mathbb{E}[\lVert K_1 - K_2 \rVert].\label{eq:exp}
\end{align}
Let us continue by working on the right hand term of line \eqref{eq:exp}. The random variable $K_1 - K_2$ has a $\text{Skellam}(q\tau, q\tau)$ distribution \cite{Skellam1946}, which has probability mass function
\begin{equation*}
p(k) = e^{-2q\tau} \cdot I_{k}(2q\tau),
\end{equation*}
where $I_{k}(\cdot)$ is the $k$th modified Bessel function of the first kind. We thus find
\begin{align}
    \label{eq:line1}\frac{1}{2}\mathbb{E}[\lVert K_1-K_2\rvert ] &= e^{-2q\tau} \cdot \sum^{\infty}_{k=0} k I_{k}(2qt)\\
    \label{eq:line2}&= q\tau e^{-2q\tau} \cdot [I_{0}(2q\tau) + I_1(2q\tau)].
\end{align}
To obtain line \eqref{eq:line1} we used the Bessel function identity $I_{-k}(\cdot) = I_{k}(\cdot)$, and to move to line \eqref{eq:line2} we use the identity $kI_{k}(z) = \frac{z}{2}[I_{k-1}(z) - I_{k+1}(z)]$ to telescope the sum in \eqref{eq:line1}.

To obtain the expected speed we integrate $f(\tau)$ the expected progress of the knowledge state between jumps \eqref{eq:prog} against the probability density of $\mathcal{T}$, and then divide by the expected length of $\mathcal{T}$. In analogy, consider finding the expected speed of the following toy process: toss a fair coin, proceed at one metre per second for two seconds if heads, proceed a two metres per second for one second if tails, keep tossing. Recalling $\mathcal{T}$ is exponentially distributed with scale parameter $(1-q)$ and making use of expressions \eqref{eq:exp} and \eqref{eq:line2}, we find
\begin{widetext}
\begin{equation}\label{eq:ints}
     \int^{\infty}_{0} (1-q)e^{-(1-q)\tau} f(\tau)\,d\tau = \int^{\infty}_{0} \tau q(1-q)e^{-(1-q)\tau} \,d\tau + \int^{\infty}_{0} \tau q(1-q)e^{-(1+q)\tau} \cdot [I_{0}(2q\tau) + I_1(2q\tau)] \,d\tau.
\end{equation}
\end{widetext}
The first integral on the right hand side of \eqref{eq:ints} is simply $q$ times the mean of an exponential distribution with scale parameter $1-q$. More surprisingly, the rightmost integral of \eqref{eq:ints} can also be evaluated explicitly using a standard Laplace transform identity for $I_{0}$. Starting with the identity
\begin{equation}\label{eq:id}
    \int^{\infty}_{0} e^{-a\tau} \cdot I_{0}(b\tau) \,d\tau = \frac{1}{\sqrt{a^2-b^2}} \quad \text{for} \quad a > |b|,
\end{equation}
we can obtain further identities which allow us to evaluate the rightmost integral of \eqref{eq:ints}.

Differentiating both sides of identity \eqref{eq:id} with respect to $a$ yields
\begin{equation}\label{eq:id1}
    \int^{\infty}_{0} \tau e^{-a\tau} \cdot I_{0}(b\tau) \,d\tau = \frac{a}{(a^2-b^2)^\frac{3}{2}} \quad \text{for} \quad a > |b|.
\end{equation}
Furthermore we know that $\frac{\partial}{\partial b} I_{0}(b\tau) = \tau I_{1}(b\tau)$ and thus differentiating both sides of \eqref{eq:id} with respect to $b$ yields
\begin{equation}\label{eq:id2}
    \int^{\infty}_{0} \tau e^{-a\tau} \cdot I_{1}(b\tau) \,d\tau = \frac{b}{(a^2-b^2)^\frac{3}{2}} \quad \text{for} \quad a > |b|.
\end{equation}
The identities \eqref{eq:id1} and \eqref{eq:id2} give us everything we need to evaluate the rightmost integral of \eqref{eq:ints} exactly. At last, to get the expected speed we divide \eqref{eq:ints} by $\mathbb{E}[\mathcal{T}] = 1/(1-q)$, perform a little algebra, and obtain equation \eqref{eq:two}.

The above argument does not obviously generalise to larger communities. Consider the case of a community of ten researchers, as plotted in Figure \ref{fig:li}: there is no obvious choice of time in which to split up the process. For instance, if we try a similar trick and split into times where some researcher has just assimilated knowledge from the most knowledgable researcher, the other researchers can get lucky and overtake both of the front running researchers via innovation actions. If such overtakes never occurred the speed for the two researcher case would agree with the simulated speed of the ten researcher case. It does not agree, see Figure \ref{fig:li}, and so such overtakes must occur. How to deal with researchers joining the front via innovation actions instead of assimilation actions is not obvious.

\section{Discussion}

We defined two IPS models which replicate a few interesting features of knowledge advancement. Most notably, we saw diminishing returns in the knowledge advancement speed as the research population grew. In the high information case, where researchers could assess who is most knowledgeable and assimilate knowledge from the front runners, the knowledge advancement speed scaled as the square root of the research population size, drawing comparison with Price's Law. For large research populations researchers had to assimilate more often to keep up with the faster moving front, and progress as a whole was thus relatively slowed. The trade-off researchers must make between innovating and assimilating was also made clear in the low information case, where researchers could not assess who was most knowledgable. A mixture of innovation and assimilation maximised the knowledge advancement speed in this case, and once again, the balance tipped more towards assimilation as research populations grew larger.

A network structure can be imposed on the researchers, such that assimilating researchers may only choose to assimilate from neighbours in the network. Our notion of `community', in which every researcher could interact with every other, is the complete network for example. The LI process was investigated on more general networks. The travelling wave methodology used to obtain estimate \eqref{eq:upest} can be readily extended to the network setting. However, even when the topology of the network is explicitly included in the analogue of equation \eqref{eq:master}, the same speed estimate results. Briefly, the analogue of equation \eqref{eq:master} follows from a mean field approximation applied to an (exact) master equation. The mean field approximation is equivalent to assuming a large metapopulation of researchers at each node in the network. For a concrete example, the travelling wave methodology applied to a network of two connected researchers would actually find the speed of the process on a (infinite) fully connected bipartite network. Less surprisingly, the knowledge advancement speed of the infinite bipartite network and the infinite complete network are identical. Finding methods which give good speed estimates (for both the HI and LI cases) on arbitrary research networks remains an open challenge.

Our intuition leads us to believe that knowledge is structured more like a tree than a line, in the sense that from a single piece of knowledge there may be multiple different pieces which follow. This is expressed nicely in Tria's model of novelties in \cite{Tria2014}, where encountering a novelty `expands the space of the possible'. Our understanding of KAPs on the line tell us something about the case knowledge in which knowledge is structured as a tree, where on each branch a similar dynamic to the line case occurs. A second complicating possibility is that unearthing a new chunk of knowledge requires understanding from multiple different domains. In this case knowledge is more naturally structured as a directed hypergraph (DH) \cite{Ausiello2017}. An edge in a DH consists of two disjoint subsets of nodes $V_{\text{in}}, \{v\}$. The first set $V_{\text{in}}$ is the input knowledge chunks required to obtain output knowledge chunk $v$. Looking further into the future, study of KAPs in which knowledge states have a more interesting topology (like a tree or directed hypergraph for example) could also prove worthwhile.

\begin{acknowledgments}
B.N. is supported by a scholarship from the EPSRC Centre for Doctoral Training in Statistical Applied Mathematics at Bath (SAMBa), under the project EP/S022945/1. M.H. was supported by the Bath Institute for Mathematical Innovation (IMI) via an undergraduate summer research placement. M.O. is partially supported by EPSRC research grant EP/X040089/1.
\end{acknowledgments}

\bibliography{apssamp}

\end{document}